\documentstyle[12pt]{article}
\title{\begin{flushright}
\normalsize{ITFA 93-05  \\ ITEP-M-1/93 \\ hep-th/9303150}
\end{flushright}
\vspace{.5cm}
Charge screening in the Higgs phase\\ of Chern-Simons electrodynamics}
\author{\normalsize{F.Alexander Bais\thanks{bais@phys.uva.nl}$\,$,
Alexei Morozov\thanks{Permanent address: ITEP 117259, Moscow,
Russia} $^,$\thanks{morozov@vxdesy.desy.de; morosov@phys.uva.nl}
$\,$,  and Mark de Wild Propitius\thanks{mdwp@phys.uva.nl}  }
\\ \\ \normalsize{Institute for Theoretical Physics}\\
\normalsize{Valckenierstraat 65,
1018XE Amsterdam, the Netherlands} }
\date{\normalsize{March 1993}}
\begin{document}
\maketitle
\begin{abstract}
Though screened at large distances, a point-like
electric charge can still participate in a long-range
electromagnetic interaction in the Higgs phase,
namely that with the Aharonov-Bohm  field
produced by a localized magnetic flux. We show that this follows
from the fact that the screening charge, induced in the presence of
a Higgs condensate, does not interact with the Aharonov-Bohm field.
The same  phenomenon occurs, if a Chern-Simons term is incorporated in
the action.
This observation provides a physical basis for the recently proposed
classification of the superselection sectors of this model in terms
of a quasi-Hopf algebra.
\end{abstract}
PACS numbers: 0365, 0530L, 1130, 1220
\newpage \noindent

The question whether charges in the Higgs phase can be measured
through  Aharonov-Bohm (AB) scattering with magnetic fluxes~\cite{ahbo},
has received quite some attention in recent literature.
These discussions mostly focussed on gedanken experiments involving
closed shells or annuli of unbroken vacuum
surrounding the charge (see for instance Ref.\ \cite{pres}), and lead to the
conclusion that Coulomb fields need not be screened inside the volume $V$
surrounded by these shells
(i.e.\ $Q_V= \int_V \partial_i F^{i0}d^Dx\neq 0$).
The point we  like to make in this letter, is  that even {\em inside} the
broken phase these charges can be measured  by means of AB scattering
with magnetic fluxes (and vice versa) due the to a
rather subtle effect, which depends on the specific details of the
Higgs mechanism. So, though the Coulomb interactions are
completely screened in the Higgs phase, i.e.\ $Q=0$, the AB interactions are
not. This effect is  not a
common feature of  any screening mechanism, but is specific
for screening by a Higgs condensate. For example, Debye
screening by a real plasma, or the
(sometimes partial) screening related to vacuum polarisation as
described by the renormalization group
(Gell-Mann-Low $\beta$-function), are not of this type.
They screen  both Coulomb, and AB interactions.
Even more peculiar is the screening, related to the Chern-Simons (CS)
term~\cite{CSE}.
It leads to complete screening
of the Coulomb -, and partial screening (by a factor 1/2)
of the AB interaction (see for instance Ref.\ \cite{wil}, and
the text below).

The model in which we discuss these
phenomena is the $2+1$ dimensional theory, governed by the  Lagrangian
\begin{eqnarray}
{\cal{L}} &=&
{\cal{L}}_{ed}(\mu) + {\cal{L}}_{Higgs} +{\cal{L}}_{matter}   \nonumber \\
&=& -\frac{1}{4}F^{\kappa\nu}F_{\kappa\nu} +\frac{\mu}{4}
\varepsilon^{\kappa\nu\lambda}F_{\kappa\nu}A_{\lambda} +
\mid{\cal{D}}_{\kappa}\Phi\mid^2 - V(\mid\Phi\mid^2) + {\cal{L}}_{matter}.
\label{Lag}
\end{eqnarray}
The first two terms (${\cal{L}}_{ed}(\mu)$) describe what is known as
CS  electrodyna\-mics~\cite{CSE}. In the next two terms (${\cal{L}}_{Higgs}$)
a complex scalar field $\Phi$ with global $U(1)$ charge $Ne$ is minimally
coupled to this gauge theory. In our conventions
${\cal{D}}_{\kappa}\Phi \equiv (\partial_{\kappa} + iNeA_{\kappa})\Phi$.
To proceed, we assume that the potential $V(\mid\Phi\mid^2)$ takes the form
\begin{eqnarray}
V(\mid\Phi\mid^2)=\frac{\lambda}{4}(\mid\Phi\mid^2-v^2)^2,
\end{eqnarray}
which spontaneously breaks the gauge symmetry $U(1)\rightarrow Z_N$,
and leads to a condensate of the charged Higgs field $\Phi$.
In this so-called Higgs phase the length scale is set by $1/M_H$, with
$M_H = \sqrt{2\lambda} \mid<\Phi>\mid= \sqrt{2\lambda}v$
the mass of the Higgs boson.
We have introduced additional charged matter fields (${\cal L}_{matter}$),
in order to be able to discuss all conceivable charge sectors in the
Higgs phase. We do not further specify these matter fields.
We only  assume that  they are very massive, so that we can discuss
the associated global $U(1)$ charges, denoted by $q$, as external.
\footnote{We take the charges $q$ to be integer multiples
of $e$, i.e.\ the model (\ref{Lag}) corresponds to compact
electrodynamics, for instance obtained from some non-abelian model after
spontaneous symmetry breaking. In the latter case
the topological mass $\mu$ would be
quantized~\cite{CSE}. In the text we assume however, that $\mu$ can take
any real value.}

Let us first recall that pure CS electrodynamics ${\cal{L}}_{ed}(\mu)$ with
 $\mu\neq 0$ is known to possess only massive particles in the
spectrum:  one-component massive photons~\cite{CSE} with  mass $\mu$.
On the other hand, the photons of pure Higgs electrodynamics
${\cal{L}}_{ed}(0) + {\cal{L}}_{Higgs}$  in the Higgs phase
are also massive. This time they are endowed
with two components, and they carry the mass $M_A = Ne \sqrt{2}\mid<\Phi>\mid=
Ne\sqrt{2}v$.
These theories differ radically  from massive
electrodynamics with  explicitly broken gauge invariance
\begin{eqnarray}                              \label{lmed}
{\cal{L}}_{med} = -\frac{1}{4}F^{\kappa\nu} F_{\kappa\nu} +
\frac{m^2}{2}A^{\nu} A_{\nu}.
\label{med}
\end{eqnarray}
Despite having only massive particles, the former models allow for the
long-range AB interaction~\cite{ahbo}, which is absent for
${\cal{L}}_{med}$. One may think of the AB
interaction in pure CS electrodynamics ${\cal{L}}_{ed}(\mu)$ with
 $\mu\neq 0$, as being mediated by certain `discrete' degrees of
freedom \cite{KM}, which are  close analogues of the by now well known
discrete states of the $c=1$ string \cite{c=1str}.

In the Higgs phase of the model ${\cal{L}}_{ed}(0) +
{\cal{L}}_{Higgs}$ the AB interaction survives, because the field,
which is acquiring mass due to the spontaneous breakdown, is actually
the gauge invariant combination
\begin{eqnarray}
\tilde{A}_{\kappa} \equiv  A_{\kappa} + \frac{1}{Ne}\partial_{\kappa}
\mbox{Im}\log<\Phi>,
\label{Atilde}
\end{eqnarray}
rather than $A_{\mu}$ itself.
Instead of (\ref{med}), we find in the Higgs
phase
\begin{eqnarray}      \label{ma}
-\frac{1}{4} F^{\kappa\nu}F_{\kappa\nu} +
\mid{\cal D}_{\kappa}\Phi\mid^2 \longrightarrow
-\frac{1}{4} F^{\kappa\nu}F_{\kappa\nu} + \frac{M_A^2}{2}
\tilde A^{\kappa}\tilde A_{\kappa}.
\label{mhed}
\end{eqnarray}
Thus $\tilde{A}$ indeed has a finite ($\sim 1/M_A$)
correlation length. This  does not immediately imply that $A$ should
also fall off exponentially. It can instead remain pure gauge. Such AB
fields, which are locally pure gauge, can be globally nontrivial
however. This is  the case
around topological defects of the Higgs condensate of the
characteristic size  $\sim 1/M_A$, corresponding to  magnetic
vortices~\cite{niels} labelled by $\pi_1(U(1))\simeq Z$.
These vortices carry a magnetic
flux $\phi$ which is quantized by the requirement that the Higgs condensate is
single-valued outside the core of the vortex, thus
\footnote{
Vortices are in fact nothing but 2-dimensional monopoles, however we
do not use the word monopole in order  to avoid confusion with the
monopoles in 3 spatial dimensions. The latter appear in fact as $instantons$ in
the
2+1 dimensional theory under consideration. These instantons cause
transitions with a change of the flux $\Delta\phi=\frac{2\pi}{e}$,
and in particular render vortices with $\phi \geq \frac{2\pi}{e}$
unstable~\cite{BDW2}. Thus instead of Eq.\ (\ref{maqu}), it
is jusitified to consider a
smaller set of fluxes: $\phi = \frac{2\pi}{Ne} \times \mbox{integer}\ {\rm
mod}\ N$ as suggested by the topological characterisation of the
fluxes. Even if there were no instantons, two fluxes $\phi_1$, and
$\phi_2$ with $\phi_1 - \phi_2 = \frac{2\pi}{e}\times \mbox{integer}$ would
be indistinguishable by AB interactions with the  charges $q$, which
are integer multiples of $e$. However, the energy of the fields inside
the core of the vortices $\phi_1$, and $\phi_2$ is different, and hence
are their masses. This may lead to another type of dynamical
instability, namely the decay of a vortex with $n$ units of flux into $n$
vortices with a single unit, as it appears in Type II
superconductors.}
\begin{eqnarray}
\phi = \frac{2\pi}{Ne}\times \mbox{integer}.
\label{maqu}
\end{eqnarray}
It is well known that the purely quantummechanical AB
interaction~\cite{ahbo,pres}
leads to non trivial elastic scattering of charges $q'$, and vortices $\phi$,
i.e.\ it gives
rise to a diffraction like effect, which is of course observable.
\footnote{If  the explicit gauge symmetry breaking term
$\frac{m^2}{2}A^2$ in~(\ref{lmed}) is introduced with $1/m \gg 1/M_A$,
then the AB field, which was pure
gauge, and carried no energy for vanishing $m$, will acquire a non-vanishing
energy. This implies that vortices will be attracted to each other
, and therefore will be confined at distances $\sim 1/m$, forming `meson'-, and
`baryon'-like structures of a vortex anti-vortex pair or
$N$ vortices respectively.}
The crucial ingredient in the corresponding
cross-sections is the phase $\exp iq'\phi$.
If there were no magnetic vortices (with their AB fields) in the Higgs phase,
electric charges $q$ would be
unobservable at $large$ distances (of course they can always be
seen in scattering processes with high energies $\geq M_A$, i.e.\ at
short distances). But even at {\em low} energies any two states with
$U(1)$ charges $q_1$, and $q_2$, such that
$q_1 - q_2 = Ne\times \mbox{integer}$
are in fact indistinguishable, because the fluxes which can be used to
distinguish them are constrained by Eq.\ (\ref{maqu}). This observation gives
rise to the low-energy classification of the superselection
sectors of this model by means of $(q,\phi)$, with
\begin{eqnarray}
q = e\times(\mbox{ integer}\ {\rm mod}\ N), \label{moe} \\
\phi = \frac{2\pi}{Ne}\times(\mbox{ integer}\ {\rm mod}\ N).
\label{abclass}
\end{eqnarray}
Together with its less-trivial generalization to the patterns of
symmetry-breaking $G \rightarrow H$ with non-abelian discrete
groups $H$   \cite{B}, this classification has been
investigated in Refs.\ \cite{pres}, and \cite{bu}-\cite{BDW2}.

What remains to be discussed is  {\em why} all the states $(q,\phi)$, as
listed in Eqs.\ (\ref{moe}), and (\ref{abclass}) can indeed be
distinguished by the AB interaction.
Let us recall that the generic AB field is a pure gauge
$A_{\mu} = g^{-1}\partial_{\mu}g$ with $g\in U(1)$.
In a non-\-simply-\-con\-nected domain,
one may have that $\oint A_{\mu}dx^{\mu} \neq 0$ along a non-\-contractable
loop, and in fact $\oint A_{\mu}dx^{\mu} = \phi$ if the contour
goes around a vortex once.
The interaction  of a point-like charge $q'$ (moving along the worldline
$\gamma$) with the
electromagnetic field can be written as
$q'\int_{\gamma} A_{\mu}dx^{\mu}$.
 In the first-quantization formalism
the phase factor arising when the charge is carried around the vortex,
therefore equals $\epsilon\{\phi,q'\} = \exp iq'\phi$.
However, in the Higgs phase the charge $q'$ is completely screened at
distances $\gg 1/M_A$. This means that it is surrounded by a `cloud'
with a total electric charge exactly equal to $-q'$. This cloud
serves to cancel the contribution of $q'$ to the Coulomb field.
{}From a physical point of view, this leads to a potential embarassment.
The problem being that the cloud will be also carried around the
vortex, and consequently it seems, that the factor
$\epsilon\{\phi,q'\}$ will be {\em unobservable}, because it should be
multiplied by $\epsilon_{\rm induced}\{\phi,-q'\} = \epsilon\{\phi,
q'\}^{-1}$. By this line of reasoning,
we are led to conclude that for a physical (i.e.\ dressed)
charge, the AB effect would be absent in the Higgs phase.

Remarkably enough,
the reasoning just given, turns out to be incorrect. While the
Coulomb interaction is exponentially damped
by the Higgs mechanism, the AB interaction is not.
In fact $\epsilon_{\rm induced} \equiv 1$, and the reason for this is
that the (induced) screening charge density is $not$ an ordinary electric
charge density, but rather the field $-M_A^2\tilde A_0$. This is clear
from  Gauss's law, implied by Eqs.\ (\ref{Lag}),
(\ref{Atilde}), and (\ref{mhed})
\begin{eqnarray}
\nabla\!\cdot\!{\bf E}({\bf x}) = q\delta({\bf x}) - M_A^2\tilde A_0({\bf x})
= q\delta({\bf x}) + q_{\rm induced}({\bf x}),
\label{Gl}
\end{eqnarray}
with $E^i \equiv F^{i0}$.
The associated induced current is
\footnote{It is clear that whenever the
modulus
of the value of the Higgs condensate is varying, as for example inside the
core of a vortex, $M_A=(Ne)^2 \sqrt{2} \chi^2$ is also varying, and can in fact
become a function of the coordinates, see Eq.\ (\ref{Lagt}) below.}
\begin{eqnarray}
j_{\rm induced}^{\kappa} = -M_A^2\tilde A^{\kappa}.
\label{incu}
\end{eqnarray}
In order for this current to interact with the AB field (produced
by some remote vortex), there should be a term in the Lagrangian of
the form\\
$-j_{\rm induced}^{\kappa}A_{\kappa}= M_A^2\tilde A^{\kappa}A_{\kappa}$.
Instead we only encounter the  term
$\frac{1}{2} M_A^2\tilde A^{\kappa}\tilde A_{\kappa}$
in Eq.\ (\ref{ma}). In other words, $q_{\rm induced}$
couples to $\tilde A$ rather than to $A$, and thus
does {\em not} feel AB fields related to remote vortices, which have
non-vanishing $A$-component but strictly vanishing $\tilde A$ at large
distances.
Thus our conclusion is that for a physical particle in the Higgs phase
the AB interaction
is sensitive to the {\em unscreened} charge only, while ordinary
interactions, like the Coulomb one, are completely screened.\footnote{
This is a special feature of charge screening in the Higgs mechanism.
The Debye screening mechanism, or the
screening related to vacuum polarisation as
described by the renormalization group,
screen  both Coulomb -, and AB interactions.
This is  inferred from the corresponding effective Lagrangians
$-\frac{1}{4}F^{\kappa\nu}F_{\kappa\nu} - \frac{m^2}{2}E^i\frac{1}{\Delta}
E^i$, and $-\frac{1}{4}F_{\kappa\nu}\frac{e_0^2}{e^2(\Delta)}F^{\kappa\nu}$
respectively.}
The phase factor associated with the AB interaction of two localized
states
$(q,\phi)$, and $(q',\phi')$ is therefore given by
\begin{eqnarray}
\epsilon\{(q,\phi),(q',\phi')\} \equiv \epsilon\{\phi,q'\}\epsilon
\{\phi',q\} = \exp i(q'\phi + q\phi').
\label{phafa}
\end{eqnarray}

Let us now turn to the situation where the CS term is present.
The masses of the electromagnetic fields are modified, and
become~\cite{pisa}
\begin{eqnarray}
M_{\pm}=\sqrt{M_A^2 + \frac{1}{2}\mu^2 \pm \mu\sqrt{
M_A^2+\frac{1}{2}\mu^2 }},
\label{mass}
\end{eqnarray}
where $+$, and $-$ stand for two different components of the photon.\footnote{
If $M_A=0$ one of the components becomes massless, but then it completely
decouples. CS electrodynamics without spontaneous breakdown of
gauge symmetry possesses only a $one$-component photon with mass $\mu$.}
The AB field carries no energy, and  is still present at large distances.
Gauss's law is affected though, it now reads
\begin{eqnarray}
\nabla\!\cdot\!{\bf E}({\bf x}) = q\delta({\bf x}) + \mu B({\bf x}) -
M_A^2\tilde A_0({\bf x}).
\label{Glm}
\end{eqnarray}
The new term, $\mu B\equiv \mu F^1 _{\;\; 2}$, implies that the
magnetic field of a vortex gene\-rates an additional electric charge
density  $\mu B({\bf x})$, the total extra charge being $\int \mu B
d^2x = \mu\phi$. This extra charge
is added to $q$, and will therefore be completely screened at distances
$\gg 1/M_-$ by the screening charge, corresponding to $-M_A^2\tilde A_0$.

The most significant effect of the CS term occurs in the phase factor
$\epsilon\{(q,\phi),(q',\phi')\}$. In fact, we may think of every
physical state $(q,\phi)$ as being composed of {\em three} parts: the
point-like global $U(1)$
charge $q$, the vortex $\phi$, and the screening charge
$q_{\rm induced} = -q -\mu\phi$ (all concentrated in the
domain of  radius $\sim 1/M_-$).
In the AB field, produced by some  remote
vortex $\phi'$,

(a) $q$ is coupled to the total flux $\phi'$,

(b) $q_{\rm induced}$ does not couple at all,

(c) and $\phi$ is coupled to $\frac{\mu}{2}\phi'$. \\
This can be seen from the form of the Lagrangian density (\ref{Lag}),
rewritten in terms of $\tilde A$
\begin{eqnarray}
-\frac{1}{4} F^{\kappa\nu} F_{\kappa\nu}  &+&
\frac{1}{4}\mu\varepsilon^{\kappa\nu\lambda} F_{\kappa\nu}A_{\lambda} +
\nonumber \\
&+& (Ne)^2\chi^2 \tilde A^{\lambda}\tilde A_{\lambda} - V(\chi^2) +
(\partial^{\kappa}\chi)(\partial_{\kappa}\chi)   + {\cal L}_{matter} ,
\label{Lagt}
\end{eqnarray}
where we substituted
$\Phi(x) = \chi(x)e^{i\sigma(x)}$ for the Higgs field,
with real valued
$\sigma(x)$ and gauge invariant $\chi(x) = \mid\Phi(x)\mid$, so
$\tilde A_{\kappa} \equiv A_{\kappa} +
\frac{1}{Ne}\partial_{\kappa}\sigma$.
The fluctuations of $\chi$ around $<\chi> = v$ describe
neutral Higgs bosons with  mass $M_H$.
Now let us assume that we are in the Higgs phase, i.e.\ $\chi$ takes
the vacuum expectation value $v$ everywhere.
We first note that $\exp(i\sigma /N)$ is not single-valued in the presence of
a vortex, it is therefore
not a well defined gauge transformation. For this reason we cannot
simply replace $A$ by $\tilde A$ in the matter term. This  implies
statement (a). The fact that $\tilde A$, rather than $A$, appears
in the interaction term with the Higgs sector, leads to the statement
(b) above. Finally, we observe that the coefficient in front of the CS
term differs by a factor of $2$ in the Lagrangian (\ref{Lagt}), and in
Gauss's law (\ref{Glm}). This difference is the origin of the factor
$\frac{1}{2}$ in (c) (see for instance Refs.\ \cite{wil,BDW2}).

{}From (a)-(c) it follows that the total effect of the AB interaction
of the two states $(q,\phi)$, and $(q',\phi')$ when $\mu \neq 0$ is
the phase factor~\cite{BDW2}
\begin{eqnarray}
\exp i\{(q + \frac{\mu}{2}\phi)\phi' + (q' + \frac{\mu}{2}\phi')\phi\} =
\exp i(q\phi' + q'\phi + \mu\phi\phi')   .
\label{phafam}
\end{eqnarray}
 This non-trivial factor arises in the amplitudes, although
 the total electric charge $Q$ of every localized
state $(q,\phi)$, if measured by the Coulomb field at distances $\geq
1/M_-$, vanishes
\begin{eqnarray}
Q \equiv q + \mu\phi + q_{\rm induced} = 0.
\label{Qva}
\end{eqnarray}

In Ref.\ \cite{BDW2}, Eq.\ (\ref{phafam}) formed the starting point
for the study of the interchange~-, and fusion properties of the states
$(q,\phi)$ in the case where $\mu\equiv pe^2/4\pi$ with $p\in Z$.
The underlying
symmetry was shown to be  the Hopf algebra $D(Z_N)$.  A
non-vanishing value for $\mu$ was reflected by a deformation of this algebra
by a non-trivial 3-cocycle on the unbroken discrete group $Z_N$.
The generalisation  to non-abelian   unbroken groups turns out to be
straightforward in this formalism

To conclude, we described a mechanism, which makes the fact that
electric -, and magnetic fields are massive both in the Higgs phase, and
in the presence of a Chern-Simons term in 2+1 dimensions, consistent with
the observability of unscreened charges at large distances
through  long-range Aharonov-Bohm interactions, which survive
in both cases. This mechanism allows for  smoothly  switching on
the Chern-Simons term in the Higgs phase, and also provides a solid basis
for the discussion of the statistical properties of the states
presented in Refs.\ \cite{BDW1}, and \cite {BDW2}.

\bigskip
A.B. and M. de W.P. thank Peter van Driel for  useful discussions.
A.M. is indebted to the Institute for Theoretical
Physics of the University of Amsterdam, and the NIKHEF for
their kind hospitality and support.


\begin{thebibliography}{12}

\bibitem{ahbo}  Y. Aharonov, and D. Bohm, Phys. Rev. {\bf 115}, 485 (1959).
\bibitem{pres}  L.~M. Krauss, and F. Wilczek, Phys. Rev. Lett. {\bf 62}, 1221
                (1989); M.G. Alford, J. March-Russell, and F. Wilczek,
 \               Nucl. Phys. {\bf B337}, 695 (1990);
                J. Preskill, and L.~M. Krauss, Nucl. Phys. {\bf B341}, 695
                (1990).
\bibitem{CSE}        J.~Schonfeld, Nucl. Phys. {\bf B185}, 157 (1981);
S. Deser, R.~Jackiw, and S.~Templeton, Ann. Phys. [N.Y.] {\bf 140}, 37 (1982).
\bibitem{wil} A. S. Goldhaber, R. MacKenzie, and F. Wilczek, Mod. Phys. Lett.
              {\bf A4}, 21 (1989).
\bibitem{KM} Y.~Kogan, and A. Morozov, Sov. Phys. JETP {\bf 61}, 1 (1985).
\bibitem{c=1str}
   I.~Klebanov, and A.~Polyakov, Mod. Phys. Lett. {\bf A6}, 3273 (1991);
   E.~Witten, Nucl. Phys. {\bf B373}, 187 (1992).
\bibitem{niels} H. B. Nielsen, and P. Olesen, Nucl. Phys. {\bf B61}, 45
(1973).
\bibitem{B}  F.~A. Bais, Nucl. Phys. {\bf B170} (FSI), 32 (1980).
\bibitem{bu}    M.~G. Alford, and J. March-Russell,
                Int. J. Mod. Phys. {\bf B5}, 2641 (1991);
                M. Bucher, Nucl. Phys. {\bf B350}, 163 (1991);
                M.~G. Alford, S. Coleman, and J. March-Russell,
                Nucl. Phys. {\bf B351}, 735 (1991);
                K. Li, Nucl. Phys. {\bf B361}, 437 (1991);
                M. Bucher, K.~Lee, and J. Preskill, Nucl. Phys. {\bf B386},
                27 (1992);
                 M.~G. Alford, and J. March-Russell, Nucl. Phys. {\bf B369},
                 276 (1992);
                 M.~G.~Alford, K. Lee,
         J. March-Russell, and J. Preskill, preprint CALT-68-1700 (1991).
\bibitem{BDW1} F. A. Bais, P. van Driel, and M. de Wild Propitius,
               Phys.~Lett.~{\bf B280}, 63 (1992).
\bibitem{BDW2} F. A. Bais, P. van Driel, and M. de Wild Propitius, prepint
               {\bf ITFA 92-8}, to be published in Nucl. Phys. {\bf B}.
\bibitem{pisa} R. D. Pisarski, and S. Rao, Phys. Rev. {\bf D32}, 2081 (1985).
\end{thebibliography}
\end{document}